\documentclass[aps, pra,epsfigure,onecolumn, superscriptaddress
]{revtex4}
\usepackage{graphicx}
\usepackage{amsmath}    
\usepackage{amsfonts}
\usepackage{ulem}
\usepackage[colorlinks, citecolor=red]{hyperref}
\usepackage{color}
\usepackage{epsfig}
\usepackage[up]{subfigure}
\usepackage{epstopdf}
\newcommand\norm[1]{\left\lVert#1\right\rVert}
\newcommand{\be}{\begin{equation}}
\newcommand{\ee}{\end{equation}}
\newcommand{\bc}{\begin{center}}
\newcommand{\ec}{\end{center}}
\newcommand{\bea}{\begin{eqnarray}}
\newcommand{\eea}{\end{eqnarray}}
\newcommand{\ba}{\begin{array}}
\newcommand{\ea}{\end{array}}

\def\bra#1{\mathinner{\langle{#1}|}}
\def\ket#1{\mathinner{|{#1}\rangle}}
\def\braket#1{\mathinner{\langle{#1}\rangle}}

\newcommand{\miniket}[1]{\vert#1\rangle}

\newcommand{\miniprod}[2]{\langle#1\vert#2\rangle}
\newcommand{\sand}[3]{\langle#1\vert#2\vert#3\rangle}

\newtheorem{theo}{Theorem}

\newenvironment{proof}[1][Proof]{\begin{trivlist}
\item[\hskip \labelsep {\bfseries #1}]}{\end{trivlist}}
\newcommand{\qed}{\nobreak \ifvmode \relax \else
      \ifdim\lastskip<1.5em \hskip-\lastskip
      \hskip1.5em plus0em minus0.5em \fi \nobreak
      \vrule height0.75em width0.5em depth0.25em\fi}

\begin{document}

\title{Localization and limit laws of a three-state alternate quantum walk on a two-dimensional lattice}

\author{Takuya Machida}
\affiliation{Japan Society for the Promotion of Science, Japan}

\author{C. M. Chandrashekar}
\affiliation{Optics and Quantum Information Group, The Institute of Mathematical
Sciences, C. I. T. Campus, Taramani, Chennai 600113, India}


\begin{abstract}
A two-dimensional discrete-time quantum walk (DTQW) can be realized by alternating a two-state DTQW in one spatial dimension followed by an evolution in the other dimension.
This was shown to reproduce a probability distribution for a certain configuration of a four-state DTQW on a two-dimensional lattice. In this work we present a three-state alternate DTQW with a parameterized coin-flip operator and show that it can produce localization that is also observed for a certain other configuration of the four-state DTQW and non-reproducible using the two-state alternate DTQW. We will present two limit theorems for the three-state alternate DTQW. One of the limit theorems describes a long-time limit of a return probability, and the other presents a convergence in distribution for the position of the walker on a rescaled space by time. We will also outline the relevance of these walks in physical systems.
\end{abstract}

\maketitle

\section{Introduction}

Quantum walks have played an important role in the area of quantum information and computation. Particularly, quantum walks have been effectively used to propose quantum algorithms~\cite{Venegas-Andraca2012} and as a tool to realize universal quantum computation~\cite{Childs2009, LovettCooperEverittTreversKendon2010}. They still continue to garner interests in simulating quantum dynamics in various physical systems and manifest the interesting phenomenon observed in real systems like photosynthesis~\cite{MohseniRebentrostLloydAspuru-Guzik2008} and edge states~\cite{AsbothObuse2013}.
Therefore, detailed studies of the quantum walks and their long-time behavior in various configurations will give a better understanding of controllable evolutions paving way for further simulating the quantum dynamics in various physically relevant systems and to engineer the quantum dynamics for the required specifications.

A discrete-time quantum walk (DTQW) on a one-dimensional position space is defined using a two-state system which is referred as a coin state. Its evolution is described using a quantum coin operation acting on a coin space followed by a position-shift operator to evolve the system coherently in superposition among different locations on a position space. The extension of the DTQW to a higher spatial dimension was successfully demonstrated by expanding the dimension of the coin  space~\cite{MackayBartlettStephensonSanders2002}. For a two-dimensional DTQW with a four-state coin space, a long-time limit distribution (theorem) was obtained by Watabe et al.~\cite{WatabeKobayashiKatoriKonno2008}. The limit theorem for the four-state DTQW was determined by a parameterized coin-flip operator which contained a Grover coin, and its limit density function was shown to feature a Dirac $\delta$-function highlighting a localized component and a continuous function with a compact support representing a diffusing component. The clear spread of probability distribution of the four-state DTQW with time indicating the absence of the Dirac $\delta$-function was realized only for a particular composition of the initial state of the walk and the coin-flip operation.

However, an extended coin space with a control over the internal states to implement the corresponding coin operations is an extremely challenging task for physical implementation of the DTQW in higher dimensions.
Therefore, an alternate scheme to implement a DTQW on a two-dimensional position space using a two-state system was introduced~\cite{ChandrashekarBanerjeeSrikanth2010, DiMcBusch2011}.
The two-state DTQW was first evolved in one spatial dimension followed by an evolution in the other dimension and this process of the alternate evolution was repeated to implement large number of steps of the walk.
This two-state alternate DTQW was shown to manifest the wide spread probability distribution for a specific configuration of a four-state DTQW on a two-dimensional position space. The long-time limit distribution describing the asymptotic behavior of the two-state alternate DTQW after a large number of steps has also been reported~\cite{DiMcMachidaBusch2011}. The absence of Dirac $\delta$-functions in the limit density function of the two-state alternate DTQW is helpful in understanding the similarities with the limit distribution function obtained for a particular configuration of the four-state DTQW. Various properties of the two-state alternate DTQW on an $N$ dimensional space was later reported~\cite{RoldanDiSilvaValcarcel2013,DiPaternostro2015}.
The effect of noise on the two-state alternate DTQW and a four-state Grover walk was also studied and the robustness of the two-state walk over the four-state walk in presence of noise was shown~\cite{ChandrashekarBusch2013}. None of these studies on the two-state alternate walks reported the presence of a Dirac $\delta$-function in the limit distribution as it was reported for some configurations of the four-state DTQW.
The absence of constant eigenvalues in the Fourier picture of the two-state alternate DTQW has been a reason for the non-localized evolutions.
However, later the existence of a time-dependent two-state alternate DTQW with periodic coin operators, which are based on the products of a Hadamard operator and phase-shift operators depending on time, was shown to manifest localization~\cite{DiPaternostro2015}.
This still leaves open the question of an alternate DTQW configuration which can manifest localization without any complex combination of coin operations.

Motivated by a study of a three-state Grover walk on a one-dimensional position space that results in localization around an initial position~\cite{InuiKonnoSegawa2005,Machida2015}, we extend the study to a three-state alternate DTQW in two-dimension in this paper.
We show that the three-state alternate DTQW localizes around the initial position.
We make an approximate analysis for probability amplitudes of the three-state DTQW.
Particularly, defining a return probability as just the probability of finding the walker at the origin, we compute its long-time limit from the approximate behavior of the walk.
On the other hand, we also focus on a rescaled space and give a convergence law for the DTQW.
The limit law shows a behavior of the walker on the rescaled space at an infinite time and the density function in that law consists of both a Dirac $\delta$-function which implies the possibility of localization and a continuous function with a compact support.
We discuss localization from the point of view of both the return probability on the non-rescaled space and a convergence in distribution on the rescaled space.

In the following section (Sec.~\ref{sec:define}) we introduce the three-state alternate DTQW on a two-dimensional square lattice and two limit theorems are given with their proofs in the sections after describing the model.
One is a long-time limit of the return probability (Sec.~\ref{sec:return_probability}) and the other is a long-time convergence in distribution on the rescaled space (Sec.~\ref{sec:convergence_in_distribution}). 
In Sec.~\ref{sec:Physical} we present an entanglement generated between the coin and position space, and between the two spatial dimensions for both forms, the three-state and the four-state DTQW.
We compare the observations and briefly discuss the possibility of physical realization of the three-state DTQW in a three-level atomic system.
In Sec.~\ref{conc}, we summarize our results and discuss future prospects.

\section{Definition of a three-state alternate quantum walk on a square lattice}
\label{sec:define}

\begin{figure}[h]
 \begin{center}
  \includegraphics[scale=0.6]{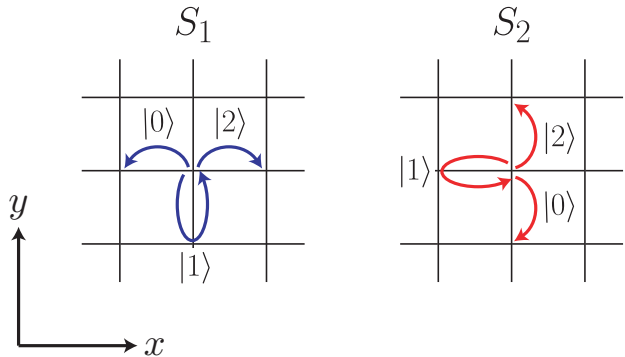}
 \end{center}
 \caption{The position-shift operator $S_1$ (resp. $S_2$) shifts the walker to the $x$-direction (resp. the $y$-direction).}
 \label{fig:position-shift_operator}
\end{figure}

In this section, we define a three-state alternate DTQW on a two-dimensional square lattice.
The position of the walker is expressed on two Hilbert spaces $\mathcal{H}_p$ and $\mathcal{H}_c$.
The Hilbert space $\mathcal{H}_c$ is spanned by an orthogonal normal basis $\left\{\ket{x,y} : x,y\in\mathbb{Z}\right\}$, where $\mathbb{Z}=\left\{0,\pm 1,\pm 2,\ldots\right\}$.
Since the Hilbert space represents the space in which the walker locates, it is called position Hilbert space.
At each vertex on the position Hilbert space $\mathcal{H}_p$, the walker can be expressed in superposition of three coin-states. Therefore, the coin Hilbert space $\mathcal{H}_c$ is spanned by an orthogonal normal basis $\left\{\ket{0}, \ket{1}, \ket{2}\right\}$. 
To compute limit laws later, we take the following orthonormal vectors: 
\begin{equation}
 \ket{0}=\begin{bmatrix}
	  1\\0\\0	  
	 \end{bmatrix},\quad
 \ket{1}=\begin{bmatrix}
	  0\\1\\0	  
	 \end{bmatrix},\quad
 \ket{2}=\begin{bmatrix}
	  0\\0\\1	  
	 \end{bmatrix}.
\end{equation}
The whole state $\ket{\Psi_t}$ of the quantum walker at time $t\,\in\left\{0,1,2,\ldots\right\}$ is described on the tensor Hilbert space $\mathcal{H}_p\otimes\mathcal{H}_c$.
The position of the walker is shifted by two position-shift operators $S_1$ and $S_2$ after the superposition is operated by a coin-flip operator $C$ as follows:
\begin{equation}
 \miniket{\Psi_{t+1}}=S_2CS_1C\ket{\Psi_t},\label{eq:time_evo}
\end{equation}
where
\begin{align}
 S_1=&\sum_{x,y\in\mathbb{Z}}\ket{x-1,y}\bra{x,y}\otimes\ket{0}\bra{0}+\ket{x,y}\bra{x,y}\otimes\ket{1}\bra{1}+\ket{x+1,y}\bra{x,y}\otimes\ket{2}\bra{2},\\
 S_2=&\sum_{x,y\in\mathbb{Z}}\ket{x,y-1}\bra{x,y}\otimes\ket{0}\bra{0}+\ket{x,y}\bra{x,y}\otimes\ket{1}\bra{1}+\ket{x,y+1}\bra{x,y}\otimes\ket{2}\bra{2},
\end{align}
and
\begin{align}
 C=\sum_{x,y\in\mathbb{Z}}\ket{x,y}\bra{x,y}\otimes &
 \Biggl(-\frac{1+c}{2}\ket{0}\bra{0}+\frac{s}{\sqrt{2}}\ket{0}\bra{1}+\frac{1-c}{2}\ket{0}\bra{2}\nonumber\\
 &+\frac{s}{\sqrt{2}}\ket{1}\bra{0}+c\ket{1}\bra{1}+\frac{s}{\sqrt{2}}\ket{1}\bra{2}+\frac{1-c}{2}\ket{2}\bra{0}+\frac{s}{\sqrt{2}}\ket{2}\bra{1}-\frac{1+c}{\sqrt{2}}\ket{2}\bra{2}\Biggr)\nonumber\\[3mm]
 =\sum_{x,y\in\mathbb{Z}}\ket{x,y}\bra{x,y}\otimes &
 \begin{bmatrix}
  -\frac{1+c}{2}& \frac{s}{\sqrt{2}}& ~~\frac{1-c}{2}\\[2mm]
  ~~\frac{s}{\sqrt{2}}& c& ~~\frac{s}{\sqrt{2}}\\[2mm]
  ~~\frac{1-c}{2}& \frac{s}{\sqrt{2}}& -\frac{1+c}{2} 
 \end{bmatrix},
\end{align}
with $c=\cos\theta$ and $s=\sin\theta\, (\theta\in [0,2\pi))$.
Since the behavior of the walker is obvious at $\theta=0,\pi$, we will not treat them. The position-shift operator $S_1$ (resp. $S_2$) plays a role of moving the walker to the $x$-direction (resp. the $y$-direction), as shown in Fig.~\ref{fig:position-shift_operator}.

\begin{figure}[h]
 \begin{center}
  \begin{minipage}{60mm}
   \includegraphics[scale=0.5]{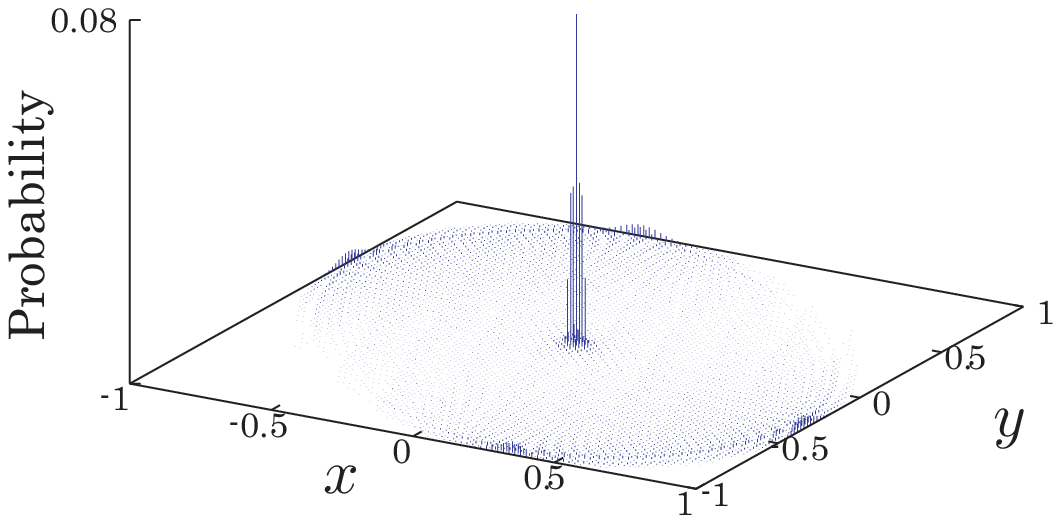}\\[2mm]
   {(a) $\alpha=\beta=\gamma=1/\sqrt{3}$}
  \end{minipage}
  \begin{minipage}{60mm}
   \includegraphics[scale=0.5]{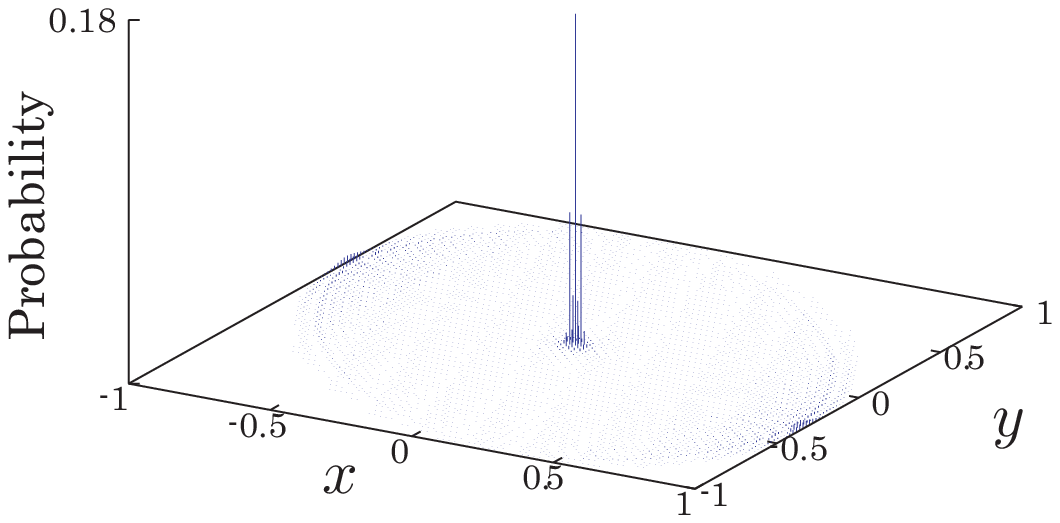}\\[2mm]
   {(b) $\alpha=\gamma=0,\, \beta=1$}
   \end{minipage}
  \caption{Probability distribution at time $50$ ($c=-1/3, s=2\sqrt{2}/3$)}
  \label{fig:distribution} 
 \end{center}
\end{figure}

\noindent
When we set $c=-1/3$ and $s=2\sqrt{2}/3$, the coin-flip operator $C$ becomes a Grover coin
\begin{equation}
 C=\sum_{x,y\in\mathbb{Z}}\ket{x,y}\bra{x,y}\otimes
  \begin{bmatrix}
  -\frac{1}{3}& \frac{2}{3}& \frac{2}{3}\\[2mm]
   \frac{2}{3}& -\frac{1}{3}& \frac{2}{3}\\[2mm]
   \frac{2}{3}& \frac{2}{3}& -\frac{1}{3}
  \end{bmatrix}.
\end{equation}
The probability, that the walker is observed at position $(x,y)\in\mathbb{Z}$, is defined by
\begin{align}
 \mathbb{P}\left[(X_t,Y_t)=(x,y)\right]=\bra{\Psi_t}\left(\ket{x,y}\bra{x,y}\otimes\sum_{j=0}^{2} \ket{j}\bra{j}\right)\ket{\Psi_t},\label{eq:prob}
\end{align}
where $(X_t,Y_t)\in\mathbb{Z}^2$ denotes the position of the walker at time $t$.
Finally we set an initial condition
\begin{equation}
 \ket {\Psi_0}=\ket{0,0}\otimes\left(\alpha\ket{0} + \beta\ket{1} + \gamma\ket{2}\right),\label{eq:initial_state}
\end{equation}
for $\alpha,\beta$, and $\gamma\in\mathbb{C}$ such that $|\alpha|^2+|\beta|^2+|\gamma|^2=1$, where $\mathbb{C}$ means the set of complex numbers.
Figure~\ref{fig:distribution} illustrates two examples of the probability distribution in Eq.~\eqref{eq:prob} and we observe localization around the origin in the pictures.


\section{Limit law of a return probability}
\label{sec:return_probability}

The study of return probabilities is one of the topics of research interest in the field of random walks as well as quantum walks.
Although there is an analytical result for a return probability of a one-dimensional DTQW as $t\to\infty$~\cite{IdeKonnoMachidaSegawa2011}, we have not had any rigorous result for two-dimensional walks. Ide et al.~\cite{IdeKonnoMachidaSegawa2011} computed a limit value of the return probability when the walker starts from a certain position, and simultaneously proved localization of the walk. 
In this section we concentrate on a return probability of the three-state alternate walk. Since the walker starts from the origin, we consider the return probability as the probability that the walker can be observed at the origin.
That is, the return probability at time $t$ is determined by the probability $\mathbb{P}[(X_t,Y_t)=(0,0)]$.
As $t\to\infty$, we obtain the following limit theorem about the return probability.

\begin{theo}
\label{th:return}
The return probability is of the form: 
 \begin{equation}
  \lim_{t\to\infty}\mathbb{P}[(X_t,Y_t)=(0,0)]=\left\{\begin{array}{ll}
						|\eta_1(\theta;\alpha,\beta,\gamma)|^2+|\eta_2(\theta;\alpha,\beta,\gamma)|^2+|\eta_1(\theta;\gamma,\beta,\alpha)|^2 & (0<\theta<\pi),\\[2mm]
						 |\eta_1(\theta-\pi;\alpha,\beta,\gamma)|^2+|\eta_2(\theta-\pi;\alpha,\beta,\gamma)|^2+|\eta_1(\theta-\pi;\gamma,\beta,\alpha)|^2 & (\pi<\theta<2\pi),	    
						      \end{array}\right.
 \end{equation}
where
 \begin{align}
  \eta_1(\theta;\alpha,\beta,\gamma)=&g_3(\theta)\alpha+\frac{1}{2}g_2(\theta)\beta+\frac{1}{2}g_1(\theta)\gamma,\\
  \eta_2(\theta;\alpha,\beta,\gamma)=&\frac{1}{2}g_2(\theta)(\alpha+\gamma)+(1-2g_3(\theta))\beta,
 \end{align}
and
 \begin{align}
  g_1(\theta)=&\frac{2\left\{\pi(1-c)^2-s(3+c^2)+4c\theta\right\}}{\pi s},\label{eq:g_1}\\
  g_2(\theta)=&\frac{\sqrt{2}\left\{\pi(1-c)+2(cs-\theta)\right\}}{\pi s},\label{eq:g_2}\\
  g_3(\theta)=&\frac{s}{\pi}.\label{eq:g_3}
 \end{align}
\end{theo}
\bigskip

\begin{proof}
We use a method based on the Fourier analysis to compute the limit of the return probability.
The Fourier analysis was introduced to quantum walks by Grimmett et al.~\cite{GrimmettJansonScudo2004}.
First, we define the Fourier transform $\ket{\hat{\Psi}_t(a,b)}\in\mathbb{C}^3\,(a,b\in [-\pi,\pi))$ of the walk at time $t$ as
\begin{equation}
 \ket{\hat{\Psi}_t(a,b)}=\sum_{x,y\in\mathbb{Z}}e^{-i(ax+by)}\ket{\psi_t(x,y)}.
\end{equation}
The amplitude at position $(x,y)\in\mathbb{Z}^2$ is extracted by using the inverse Fourier transform
\begin{equation}
 \ket{\psi_t(x,y)}=\int_{-\pi}^\pi\frac{da}{2\pi} \int_{-\pi}^\pi\frac{db}{2\pi}\,\, e^{i(ax+by)}\ket{\hat{\Psi}_t(a,b)},
\end{equation}
Equation~\eqref{eq:time_evo} leads us to the time-evolution of the Fourier transform
\begin{align}
 \ket{\hat\Psi_t(a,b)}=&R(b)\hat{C}R(a)\hat{C}\ket{\hat\Psi_t(a,b)},\label{eq:rec}
\end{align}
where
\begin{equation}
 \hat{C}=\begin{bmatrix}
	  -\frac{1+c}{2}& \frac{s}{\sqrt{2}}& ~~\frac{1-c}{2}\\[2mm]
	  ~~\frac{s}{\sqrt{2}}& c& ~~\frac{s}{\sqrt{2}}\\[2mm]
	  ~~\frac{1-c}{2}& \frac{s}{\sqrt{2}}& -\frac{1+c}{2} 
	 \end{bmatrix},\quad
 R(k)=\begin{bmatrix}
	e^{ik} & 0 & ~0\\
	0 & 1 & ~0\\
	0 & 0 & ~e^{-ik}    
       \end{bmatrix}.
\end{equation}
From Eq.~\eqref{eq:rec}, the Fourier transform at time $t$ becomes $\ket{\hat{\Psi}_t(a,b)}=\left(R(b)\hat{C}R(a)\hat{C}\right)^t\ket{\hat{\Psi}_0(a,b)}$.
We express the eigenvalues $\lambda_j(a,b)\,(j=1,2,3)$ of the unitary matrix $R(b)\hat{C}R(a)\hat{C}$ as follows:
\begin{equation}
 \lambda_j(a,b)=e^{i\nu_j(a,b)}\quad(j=1,2,3),
\end{equation}
with
\begin{align}
 \nu_1(a,b)=&0,\\[1mm]
 \nu_2(a,b)=&2\arccos\left(\frac{1+c}{2}\cos\left(\frac{a+b}{2}\right)+\frac{1-c}{2}\cos\left(\frac{a-b}{2}\right)\right),\\[1mm]
 \nu_3(a,b)=&-2\arccos\left(\frac{1+c}{2}\cos\left(\frac{a+b}{2}\right)+\frac{1-c}{2}\cos\left(\frac{a-b}{2}\right)\right).
\end{align}
The components of the normalized eigenvector $\ket{v_j(a,b)}\,(j=1,2,3)$ associated to the eigenvalue $\lambda_j(a,b)$ are given by
\begin{align}
 \braket{0|v_j(a,b)}=&-\frac{s(1-e^{-ia})}{\sqrt{N_j(a,b)}}\Bigl[e^{ib}\left\{(1+c)e^{ia}+1-c\right\}\lambda_j(a,b)-\left\{(1-c)e^{ia}+1+c\right\}\Bigr],\\[1mm]
 \braket{1|v_j(a,b)}=&\frac{2\sqrt{2}}{\sqrt{N_j(a,b)}}\Bigl[\lambda_j(a,b)^2-\left\{(1+c^2)\cos a\cos b-2c\sin a\sin b+s^2\cos b\right\}\lambda_j(a,b)+s^2\cos a+c^2\Bigr],\\[1mm]
 \braket{2|v_j(a,b)}=&\frac{s(1-e^{-ia})}{\sqrt{N_j(a,b)}}\Bigl[e^{-ib}\left\{(1-c)e^{ia}+1+c\right\}\lambda_j(a,b)-\left\{(1+c)e^{ia}+1-c\right\}\Bigr],
\end{align}
where $N_j(a,b)\,(j=1,2,3)$ is a normalized factor.

Here, we define a function
\begin{equation}
 F(x,y)=\int_{-\pi}^\pi \frac{da}{2\pi} \int_{-\pi}^\pi \frac{db}{2\pi}\,\,\frac{e^{i(ax+by)}}{16\left\{1-\left(\frac{1+c}{2}\cos\left(\frac{a+b}{2}\right)+\frac{1-c}{2}\cos\left(\frac{a-b}{2}\right)\right)^2\right\}}\quad(x,y\in\mathbb{Z}).\label{eq:definition_g}
\end{equation}
We use this function later to express the asymptotic behavior of the probability amplitude $\ket{\psi_t(x,y)}$ after many steps.
By using the residue theorem in Eq.~\eqref{eq:definition_g}, we get an integral representation of the function $F(x,y)$
\begin{align}
 F(x,y)=&\frac{1}{8\pi(1-c)}\int_{0}^{\frac{\pi}{2}} \cos((x+y)k)\left\{\frac{\left(w_1(k)-\sqrt{w_1(k)^2-1}\right)^{|x-y|}}{\sqrt{w_1(k)^2-1}}+\frac{\left(w_2(k)+\sqrt{w_2(k)^2-1}\right)^{|x-y|}}{\sqrt{w_2(k)^2-1}}\right\}dk,\label{eq:single_value_integral}
\end{align}
where
\begin{equation}
 w_1(k)=\frac{2-(1+c)\cos k}{1-c},\quad w_2(k)=\frac{-2-(1+c)\cos k}{1-c}. 
\end{equation}
Again, we get a long-time asymptotic behavior of the amplitude at position $(x,y)\in\mathbb{Z}^2$
\begin{align}
 \ket{\psi_t(x,y)}= & \int_{-\pi}^\pi \frac{da}{2\pi} \int_{-\pi}^\pi \frac{db}{2\pi}\,\,\sum_{j=1}^3 e^{i(ax+by)}\lambda_j(a,b)^t \braket{v_j(a,b)|\hat\Psi_0(a,b)}\ket{v_j(a,b)}\nonumber\\
 \sim & \int_{-\pi}^\pi \frac{da}{2\pi} \int_{-\pi}^\pi \frac{db}{2\pi}\,\,e^{i(ax+by)}\braket{v_1(a,b)|\hat\Psi_0(a,b)}\ket{v_1(a,b)}\quad (t\to\infty),
 \label{eq:asymptotic_behavior}
\end{align}
where $h_1(t)\sim h_2(t)\,(t\to\infty)$ means $\lim_{t\to\infty}h_1(t)/h_2(t)=1$.
The Riemann-Lebesgue lemma has been used in Eq.~\eqref{eq:asymptotic_behavior}.
We are also allowed to employ another form of the eigenvector
\begin{equation}
 \ket{v_1(k)}=\frac{1}{\sqrt{\tilde{N}_1(k)}}\begin{bmatrix}
				      \sqrt{2}s\Bigl(e^{i(a+b)/2}-e^{-i(a-b)/2}\Bigr)\\[2mm]
				      2i\Bigl\{(1+c)\sin((a+b)/2)-(1-c)\sin((a-b)/2)\Bigr\}\\[2mm]
				      -\sqrt{2}s\Bigl(e^{-i(a+b)/2}-e^{i(a-b)/2}\Bigr)
				     \end{bmatrix},\label{eq:v_1(k)}
\end{equation}
with
\begin{equation}
 \tilde{N}_1(a,b)=16\left\{1-\left(\frac{1+c}{2}\cos\left(\frac{a+b}{2}\right)+\frac{1-c}{2}\cos\left(\frac{a-b}{2}\right)\right)^2\right\}.
\end{equation}
Estimating Eq.~\eqref{eq:asymptotic_behavior} with Eq.~\eqref{eq:v_1(k)}, we have an expression of the asymptotic behavior of the probability amplitude $\ket{\psi_t(x,y)}$ at a large enough time,
\begin{align}
 \braket{0|\psi_t(x,y)}\sim \sqrt{2}s & \biggl\{-A(\alpha,\beta)F(x-1,y)-B(\gamma,\beta)F(x-1,y+1)+\Bigl(A(\alpha,\beta)+B(\alpha,\beta)\Bigr)F(x,y)\nonumber\\
 &+\Bigl(A(\gamma,\beta)+B(\gamma,\beta)\Bigr)F(x,y+1)-B(\alpha,\beta)F(x+1,y)-A(\gamma,\beta)F(x+1,y+1)\biggr\},\label{eq:1st-state}\\[1mm]
 \braket{1|\psi_t(x,y)}\sim (1+c) & \Bigl(-A(\alpha,\beta)F(x-1,y-1)-B(\gamma,\beta)F(x-1,y)+B(\alpha,\beta)F(x,y-1)+A(\alpha+\gamma,2\beta)F(x,y)\nonumber\\
 &+B(\gamma,\beta)F(x,y+1)-B(\alpha,\beta)F(x+1,y)-A(\gamma,\beta)F(x+1,y+1)\Bigr)\nonumber\\
 -(1-c)& \Bigl(-A(\alpha,\beta)F(x-1,y)-B(\gamma,\beta)F(x-1,y+1)+A(\alpha,\beta)F(x,y-1)+B(\alpha+\gamma,2\beta)F(x,y)\nonumber\\
 &+A(\gamma,\beta)F(x,y+1)-B(\alpha,\beta)F(x+1,y-1)-A(\gamma,\beta)F(x+1,y)\Bigr),\label{eq:2nd-state}\\[1mm]
 \braket{2|\psi_t(x,y)}\sim \sqrt{2}s & \biggl\{-A(\alpha,\beta)F(x-1,y-1)-B(\gamma,\beta)F(x-1,y)+\Bigl(A(\alpha,\beta)+B(\alpha,\beta)\Bigr)F(x,y-1)\nonumber\\
 &+\Bigl(A(\gamma,\beta)+B(\gamma,\beta)\Bigr)F(x,y)-B(\alpha,\beta)F(x+1,y-1)-A(\gamma,\beta)F(x+1,y)\biggr\},\label{eq:3rd-state}
\end{align}
where
\begin{equation}
 A(z_1,z_2)=\sqrt{2}sz_1+(1+c)z_2,\quad B(z_1,z_2)=\sqrt{2}sz_1-(1-c)z_2.
\end{equation}
Computing the long-time asymptotic behavior of the amplitude at the origin, we obtain
 \begin{align}
  \ket{\psi_t(0,0)}\sim&\left\{\begin{array}{ll}
			 \begin{bmatrix}
			  g_3(\theta)\alpha+\frac{1}{2}g_2(\theta)\beta+\frac{1}{2}g_1(\theta)\gamma\\[1mm]
			  \frac{1}{2}g_2(\theta)(\alpha+\gamma)+(1-2g_3(\theta))\beta\\[1mm]
			  \frac{1}{2}g_1(\theta)\alpha+\frac{1}{2}g_2(\theta)\beta+g_3(\theta)\gamma			  
			 \end{bmatrix} & (0<\theta<\pi),\\[8mm]
			 \begin{bmatrix}
			  g_3(\theta-\pi)\alpha+\frac{1}{2}g_2(\theta-\pi)\beta+\frac{1}{2}g_1(\theta-\pi)\gamma\\[1mm]
			  \frac{1}{2}g_2(\theta-\pi)(\alpha+\gamma)+(1-2g_3(\theta-\pi))\beta\\[1mm]
			  \frac{1}{2}g_1(\theta-\pi)\alpha+\frac{1}{2}g_2(\theta-\pi)\beta+g_3(\theta-\pi)\gamma
			 \end{bmatrix} & (\pi<\theta<2\pi),
			       \end{array}\right.
  \label{eq:asymptotic_behavior_at_origin}
 \end{align}
recalling the functions $g_j(\theta)\,(j=1,2,3)$ in Eqs.~\eqref{eq:g_1},\,\eqref{eq:g_2}, and \eqref{eq:g_3}.
The limit of the return probability follows from Eq.~\eqref{eq:asymptotic_behavior_at_origin}.
\qed
\end{proof}

\begin{figure}[h]
 \begin{center}
  \begin{minipage}{60mm}
   \includegraphics[scale=0.5]{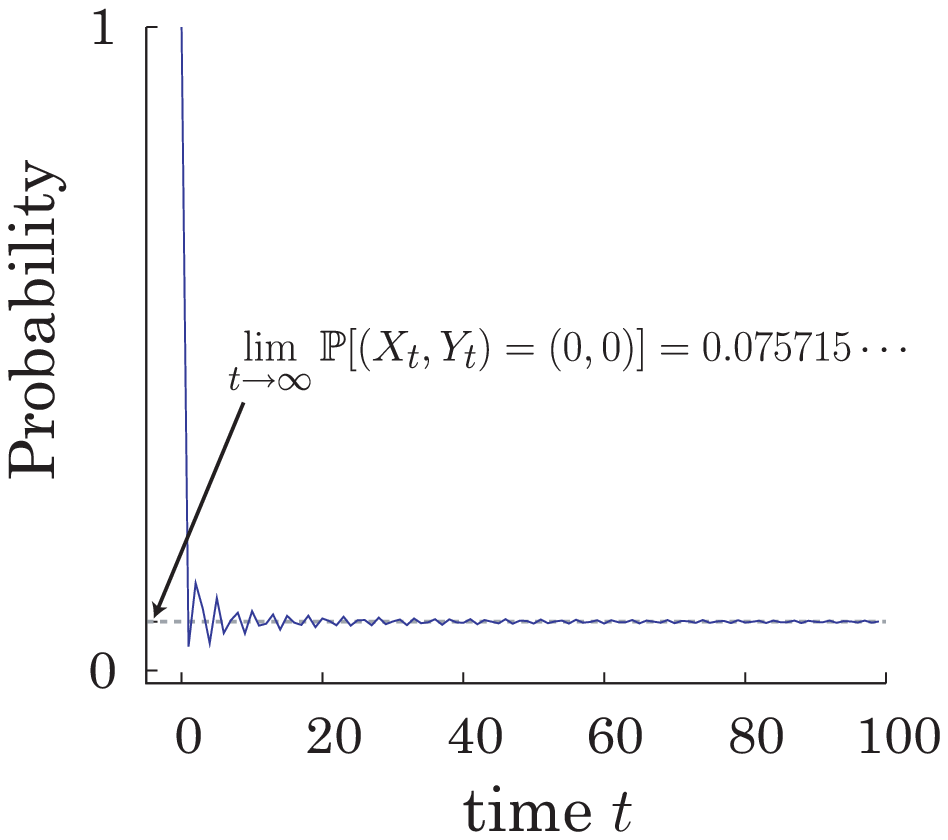}\\[2mm]
   {(a) $\alpha=\beta=\gamma=1/\sqrt{3}$}
  \end{minipage}
  \begin{minipage}{60mm}
   \includegraphics[scale=0.5]{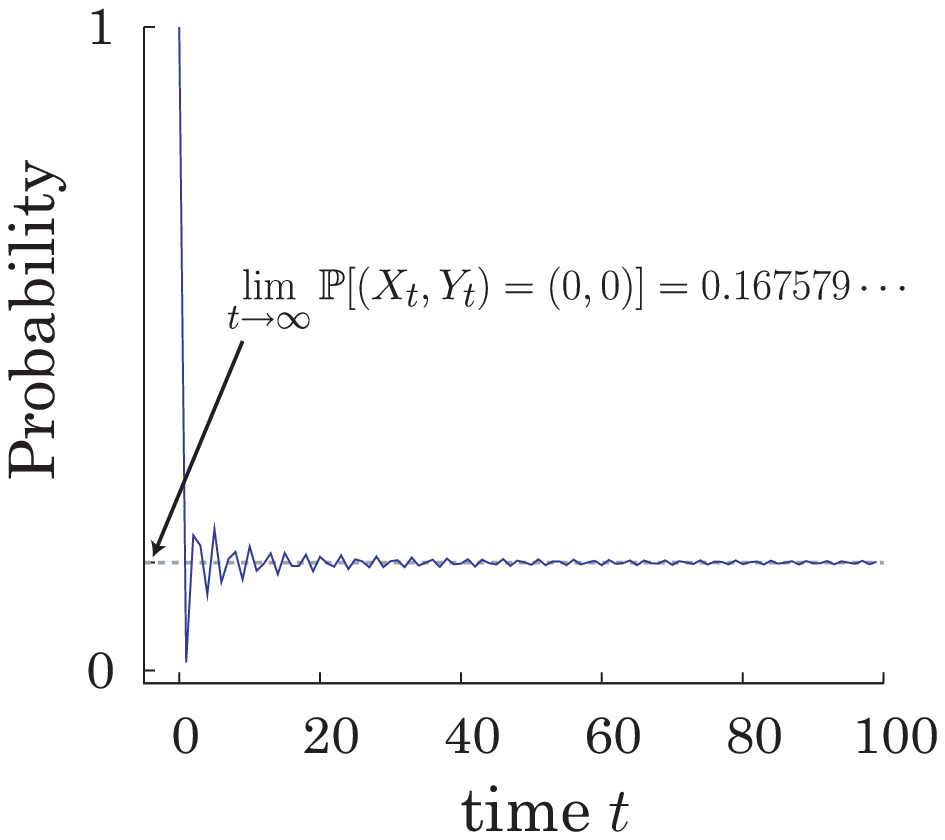}\\[2mm]
   {(b) $\alpha=\gamma=0,\, \beta=1$}
   \end{minipage}
  \caption{Given the parameter $\theta$ which satisfies $\cos\theta=-1/3$ and $\sin\theta=2\sqrt{2}/3$, the left (resp. right) figure shows how the return probability $\mathbb{P}[(X_t,Y_t)=(0,0)]$ depends on time $t$ in the case of $\alpha=\beta=\gamma=1/\sqrt{3}$ (resp. $\alpha=\gamma=0,\, \beta=1$).}
  \label{fig:return_probability_time} 
 \end{center}
\end{figure}

\begin{figure}[h]
 \begin{center}
  \begin{minipage}{60mm}
   \includegraphics[scale=0.5]{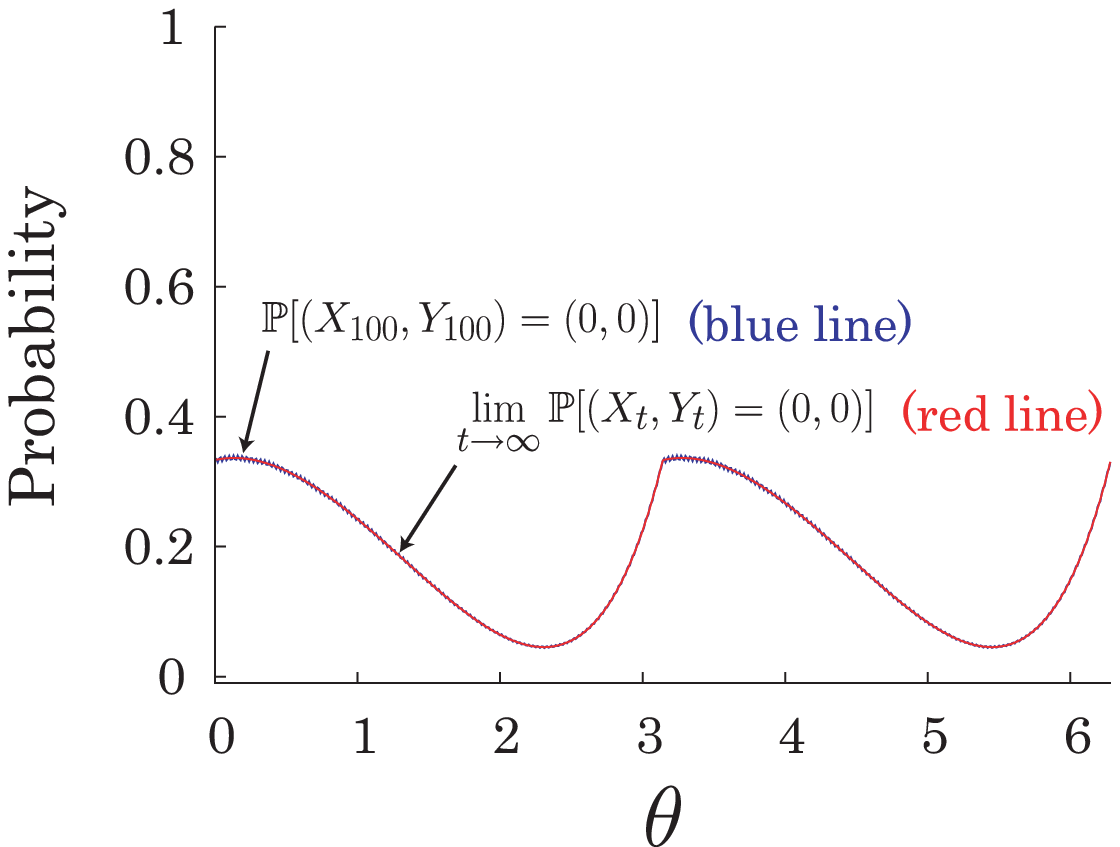}\\[2mm]
   {(a) $\alpha=\beta=\gamma=1/\sqrt{3}$}
  \end{minipage}
  \begin{minipage}{60mm}
   \includegraphics[scale=0.5]{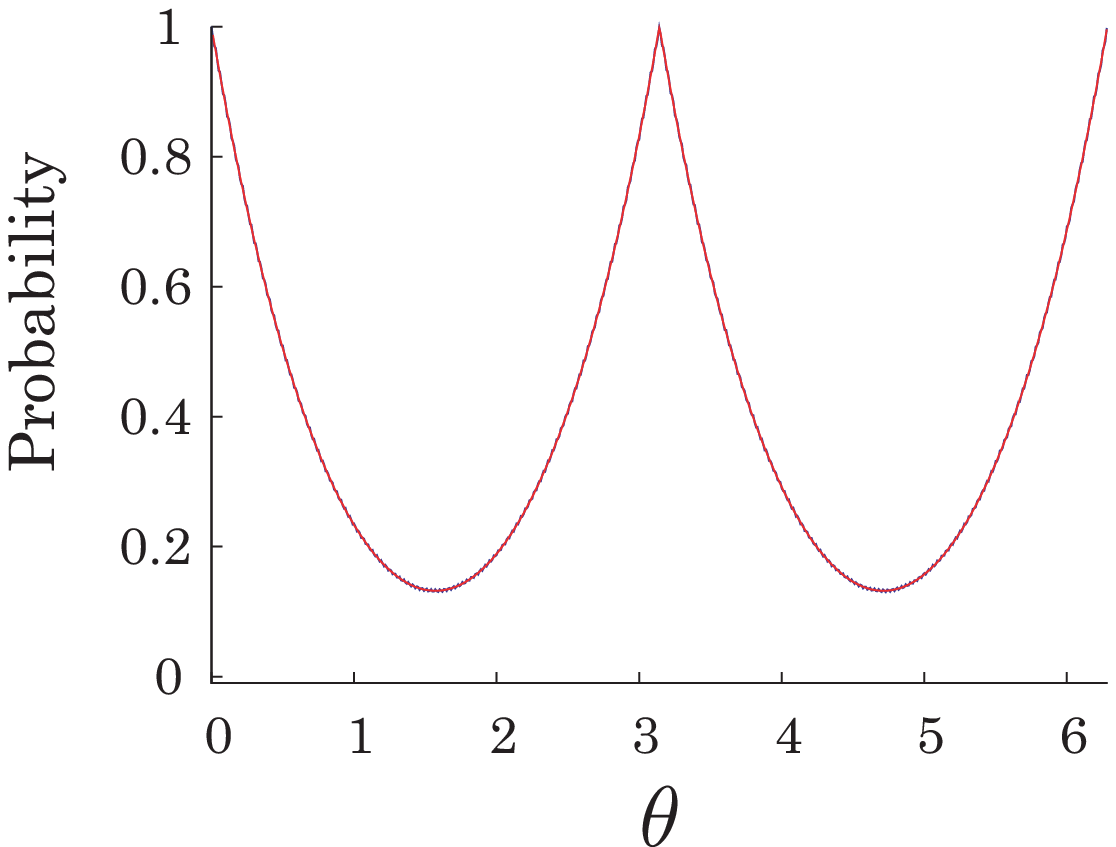}\\[2mm]
   {(b) $\alpha=\gamma=0,\, \beta=1$}
   \end{minipage}
  \caption{Return probability at time $100$ (blue line) and the limit (red line). We observe the likelihood that the walker is found at the origin at time 100, and its limit as $t\to\infty$. The positive value of the limit implies localization of the walker at the origin.}
  \label{fig:return_probability_theta} 
 \end{center}
\end{figure}

In Fig.~\ref{fig:return_probability_time} we show the return probability for two different initial states of the walk. We can see that the probability converges to the limit as time $t$ goes up. In Fig.~\ref{fig:return_probability_theta} we show the probability at time $100$ and at the limit with regard to the parameter $\theta$, which determines the coin-flip operator $C$.

\section{Convergence in distribution on a rescaled space by time}
\label{sec:convergence_in_distribution}

In the previous section, we concentrated on the probability $\mathbb{P}[(X_t,Y_t)=(x,y)]$ and computed the long-time limit of the return probability $\mathbb{P}[(X_t,Y_t)=(0,0)]$. In this section we will present a convergence theorem on a rescaled space by time. This theorem shows us the overall behavior of the walker after many steps.

\begin{theo}
\label{th:convergence_in_distribution}
The three-state alternate DTQW starting from the origin has a convergence law
\begin{equation}
\lim_{t\to\infty}\mathbb{P}\left(\frac{X_t}{t}\leq x,\,\frac{Y_t}{t}\leq y\right)=\int_{-\infty}^x du \int_{-\infty}^y dv\,\left\{\,\Delta(\theta;\alpha,\beta,\gamma)\delta_o(u,v) + \frac{\xi(u,v;\alpha,\beta,\gamma)}{2\pi^2(1-u^2)(1-v^2)}I_{\mathcal{D}}(u,v)\right\},
\end{equation}
where $\delta_o(x,y)$ denotes a Dirac $\delta$-function at the origin and
\begin{align}
 \Delta(\theta;\alpha,\beta,\gamma)=&\left\{\begin{array}{ll}
				     |\beta|^2+\Re(\alpha\overline{\gamma})g_1(\theta)+\Re((\alpha+\gamma)\overline{\beta})g_2(\theta)+(1-3|\beta|^2)g_3(\theta) & (0<\theta<\pi),\\[2mm]
					    |\beta|^2+\Re(\alpha\overline{\gamma})g_1(\theta-\pi)+\Re((\alpha+\gamma)\overline{\beta})g_2(\theta-\pi)+(1-3|\beta|^2)g_3(\theta-\pi) & (\pi<\theta<2\pi),
					   \end{array}\right.\label{eq:delta}\\
 \xi(x,y;\alpha,\beta,\gamma)= & (1-y)^2|\alpha|^2+2(1-y^2)|\beta|^2+(1+y)^2|\gamma|^2+\frac{2\sqrt{2}(x-cy)(1-y)}{s}\Re(\alpha\overline{\beta})\nonumber\\
 & -\frac{2\sqrt{2}(x-cy)(1+y)}{s}\Re(\beta\overline{\gamma})+\frac{2\left\{s^2-2x^2-(1+c^2)y^2+4cxy\right\}}{s^2}\Re(\alpha\overline{\gamma}),\label{eq:xi}\\
 \mathcal{D}=&\left\{(x,y)\in\mathbb{R}^2\,\Bigl|\,\frac{(x+y)^2}{2(1+c)}+\frac{(x-y)^2}{2(1-c)}<1\right\}.\label{eq:domain}
\end{align}
\end{theo}
\bigskip

\begin{proof}
Using the eigenvalues $\lambda_j(a,b)$ and the normalized eigenvectors $\miniket{v_j(a,b)}\,(j=1,2,3)$ of the matrix $R(b)\hat{C}R(a)\hat{C}$, the $(r_1, r_2)$-th joint moments ($r_1, r_2=0,1,2,\ldots$) of the random variable $(X_t,Y_t)$ can be expressed as
\begin{align}
 \mathbb{E}(X_t^{r_1}Y_t^{r_2})=&\sum_{(x,y)\in\mathbb{Z}^2} x^{r_1}y^{r_2}\mathbb{P}[(X_t,Y_t)=(x,y)]\nonumber\\
 =&\int_{-\pi}^\pi\frac{da}{2\pi}\int_{-\pi}^\pi\frac{db}{2\pi}\sand{\hat\Psi_t(a,b)}{D_a^{r_1}D_b^{r_2}}{\hat\Psi_t(a,b)}\nonumber\\\
 =&(t)_{r_1+r_2}\int_{-\pi}^\pi\frac{da}{2\pi}\int_{-\pi}^\pi\frac{db}{2\pi}\sum_{j=1}^3\left(\frac{D_a\lambda_j(a,b)}{\lambda_j(a,b)}\right)^{r_1}\left(\frac{D_b\lambda_j(a,b)}{\lambda_j(a,b)}\right)^{r_2}\Bigl|\miniprod{v_j(a,b)}{\hat\Psi_0(a,b)}\Bigr|^2\nonumber\\\
 &+O(t^{r_1+r_2-1}),
\end{align}
with $D_a=i(\partial{}/\partial{a})$, $D_b=i(\partial{}/\partial{b})$, and $(t)_r=t(t-1)\times\cdot\cdot\cdot\times(t-r+1)$, where $\mathbb{E}(X)$ denotes the expected value of a random variable $X$.
Obviously, we have $D_a\lambda_1(a,b)/\lambda_1(a,b)=D_b\lambda_1(a,b)/\lambda_1(a,b)=0$ because of the constant eigenvalue $\lambda_1(a,b)=1$.
The eigenvalue $\lambda_1(a,b)$, hence, causes a Dirac $\delta$-function at the origin in the limit distribution which we are trying to prove.
That means there is a possibility that localization occurs at the origin on the rescaled space $(X_t/t,Y_t/t)$.
The measure of localization generally depends on the initial condition of the walker, which is characterized by the parameter $\alpha,\beta$, and $\gamma$ in this study.
The dependence on the initial condition is expressed as the coefficient of the Dirac $\delta$-function.
On the other hand, since we compute
\begin{align}
 \frac{D_a\lambda_j(a,b)}{\lambda_j(a,b)}=-(-1)^j\frac{(1+c)\sin\left(\frac{a+b}{2}\right)+(1-c)\sin\left(\frac{a-b}{2}\right)}{\sqrt{4-\left\{(1+c)\cos\left(\frac{a+b}{2}\right)+(1-c)\cos\left(\frac{a-b}{2}\right)\right\}^2}} \quad (j=2,3),\\
 \frac{D_b\lambda_j(a,b)}{\lambda_j(a,b)}=-(-1)^j\frac{(1+c)\sin\left(\frac{a+b}{2}\right)-(1-c)\sin\left(\frac{a-b}{2}\right)}{\sqrt{4-\left\{(1+c)\cos\left(\frac{a+b}{2}\right)+(1-c)\cos\left(\frac{a-b}{2}\right)\right\}^2}} \quad (j=2,3),
\end{align}
the eigenvalues $\lambda_j(a,b)\,(j=2,3)$ give the continuous part in the limit density function.
For the joint moments of the rescaled position $(X_t/t, Y_t/t)$, by putting $D_a\lambda_j(a,b)/\lambda_j(a,b)=x$ and $D_b\lambda_j(a,b)/\lambda_j(a,b)=y$ after $t\to\infty$, we get a convergence theorem
\begin{align}
 \lim_{t\rightarrow\infty}\mathbb{E}\left[\left(\frac{X_t}{t}\right)^{r_1}\left(\frac{Y_t}{t}\right)^{r_2}\right]
 =&\int_{-\pi}^\pi\frac{da}{2\pi}\int_{-\pi}^\pi\frac{db}{2\pi}\sum_{j=1}^3\left(\frac{D_a\lambda_j(a,b)}{\lambda_j(a,b)}\right)^{r_1}\left(\frac{D_b\lambda_j(a,b)}{\lambda_j(a,b)}\right)^{r_2}\Bigl|\miniprod{v_j(a,b)}{\hat\Psi_0(a,b)}\Bigr|^2\nonumber\\\
 =&\int_{-\infty}^\infty dx\int_{-\infty}^\infty dy\,x^{r_1}y^{r_2} \left\{\,\Delta(\theta;\alpha,\beta,\gamma)\delta_o(x,y) + \frac{\xi(x,y;\alpha,\beta,\gamma)}{2\pi^2(1-x^2)(1-y^2)}I_{\mathcal{D}}(x,y)\right\},
\label{limE}
\end{align}
with Eqs.~\eqref{eq:delta},\,\eqref{eq:xi} and \eqref{eq:domain}.
Equation~\eqref{limE} guarantees Theorem~\ref{th:convergence_in_distribution}.
\qed
\end{proof}
\begin{figure}[h]
 \begin{center}
  \begin{minipage}{70mm}
   \includegraphics[scale=0.60]{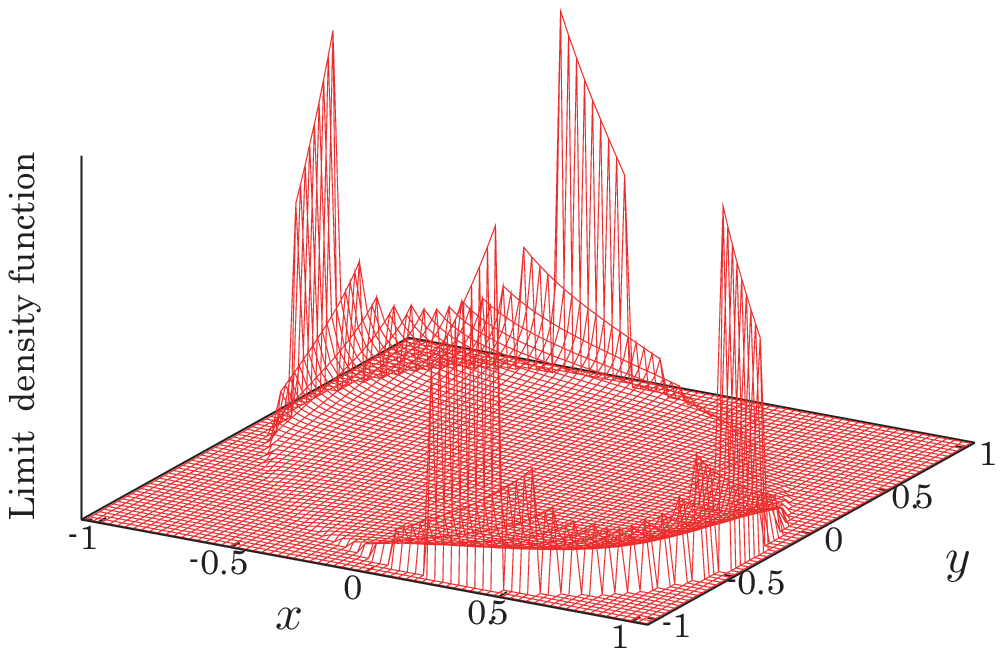}\\[2mm]
   {(a) $\alpha=\beta=\gamma=1/\sqrt{3}$}
  \end{minipage}
  \begin{minipage}{70mm}
   \includegraphics[scale=0.60]{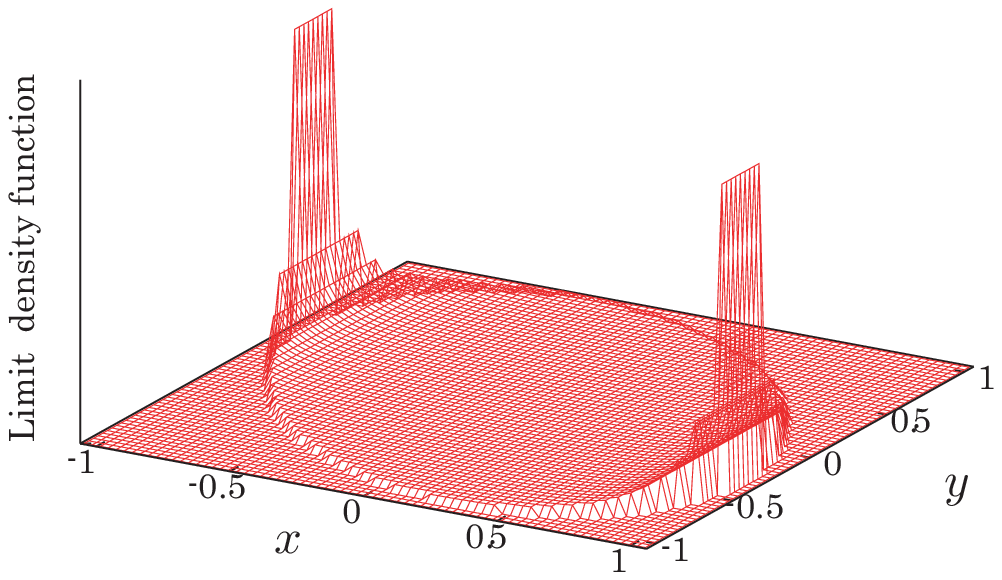}\\[2mm]
   {(b) $\alpha=\gamma=0,\, \beta=1$}
   \end{minipage}
  \caption{Continuous part of the limit density function ($c=-1/3, s=2\sqrt{2}/3$)}
  \label{fig:limit_density}
 \end{center}
\end{figure}

In Fig.~\ref{fig:limit_density} we show the two examples of the continuous part in the limit density function for a representative initial state when the walker is evolved using the Grover coin.

\section{Entanglement Generation in the three-state alternate walk and a four-state walk}
\label{sec:Physical}

From the earlier result of a four-state DTQW~\cite{WatabeKobayashiKatoriKonno2008} and from the theorems presented in Secs.~\ref{sec:return_probability} and ~\ref{sec:convergence_in_distribution}, we clearly see the presence of a Dirac $\delta$-function resulting in localization around the origin.
These localized components are also accompanied by a diffusing component as a continuous part of the limit density function.
In spite of the similarities, different dimensions of the coin space used for both the walks result in a final state which are very different from one another.
To make a fair comparison of the two kinds of localized walks, we can use the comparison of a spatial entanglement between the two-spatial dimension ($x-y$ spatial entanglement) which is common to both after tracing out the coin space from the final state~\cite{DiMcBusch2011, ChandrashekarBusch2013}. 

\begin{figure}[h]
 \begin{center}
  \begin{minipage}{70mm}
   \includegraphics[scale=0.32]{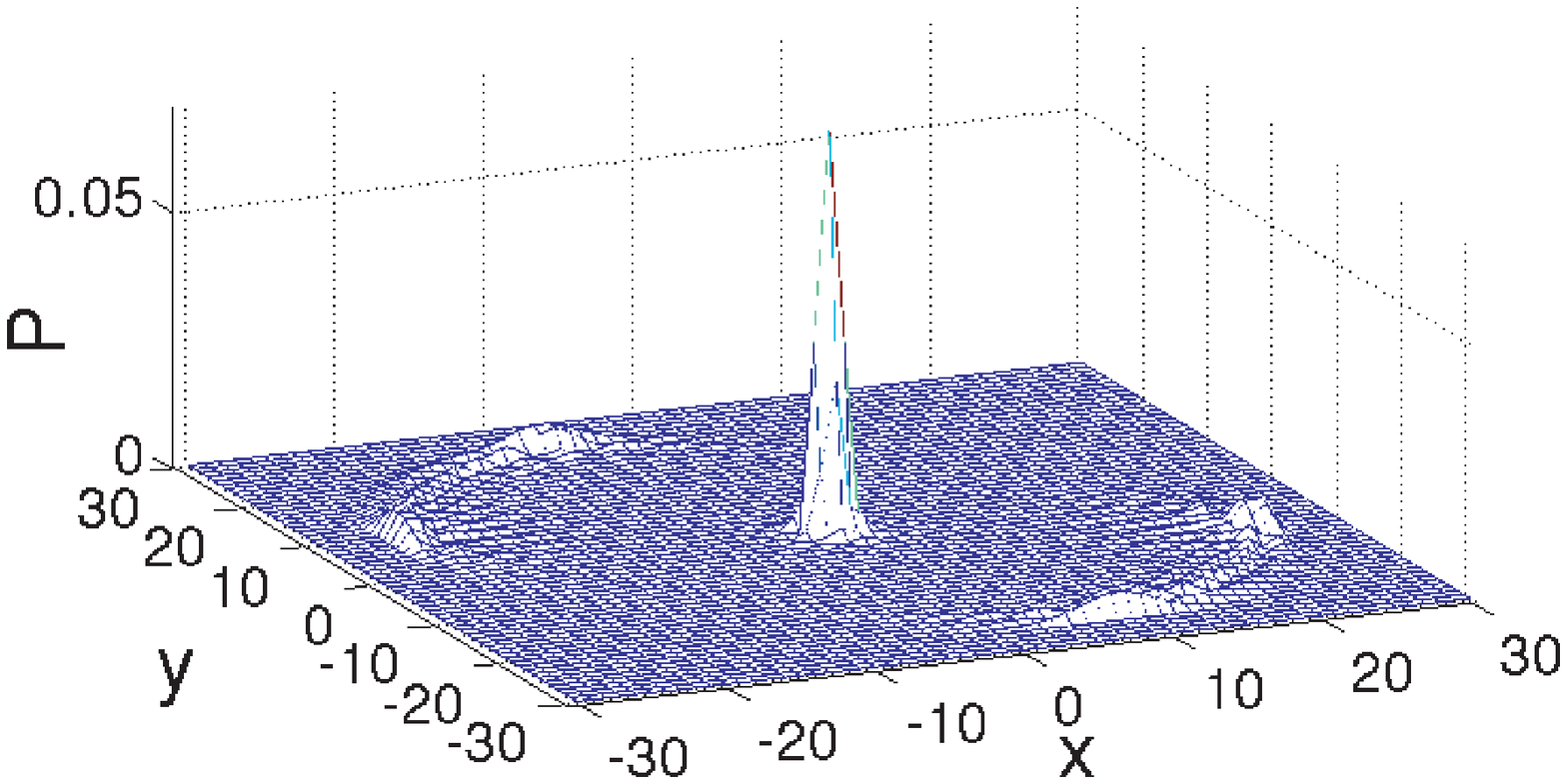}\\[2mm]
   {(a) Localization in the three-state alternate DTQW}
  \end{minipage}
  \begin{minipage}{70mm}
   \includegraphics[scale=0.32]{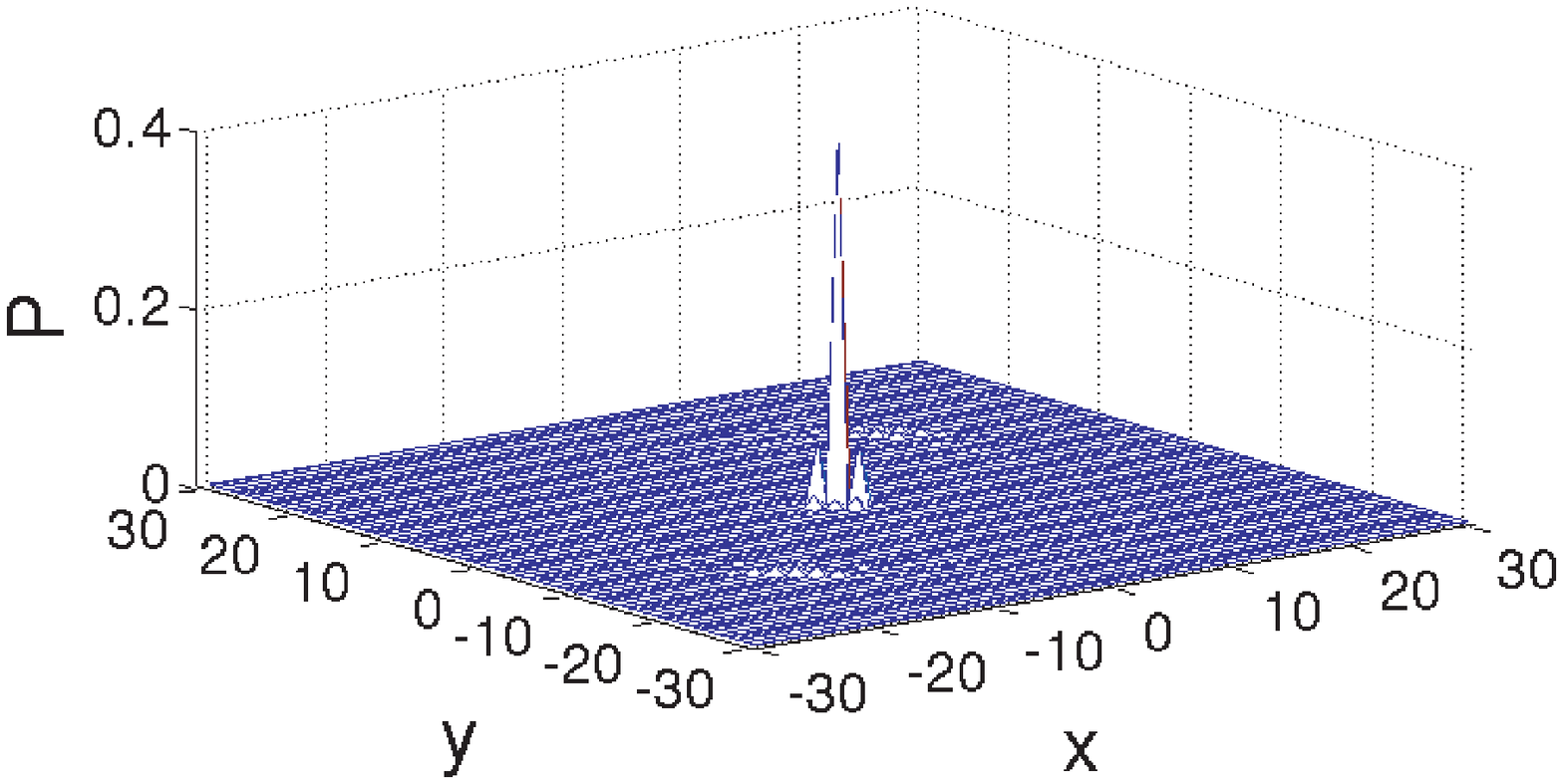}\\[2mm]
   {(b) Localization in a four-state DTQW}
   \end{minipage}
  \caption{Probability distribution at time $t=30$ (a) Three-state alternate walk with the parameters $\theta=\pi/2$, $\alpha=0$, $\beta=1$, and $\gamma=0$ which give the coefficient $\Delta=(1-2/\pi)$. (b) Four-state walk with coin parameters $p=q=1/2$ and initial state parameters $q_1=1/\sqrt{2}$, $q_2=1/\sqrt{2}$, $q_3=0$, and $q_4=0$ for the walk in Ref.~\cite{WatabeKobayashiKatoriKonno2008} which give the coefficient $\Delta=(1-2/\pi)$.}
    \label{fig:Probdistribution} 
 \end{center}
\end{figure}

In Fig.~\ref{fig:Probdistribution} we show the probability distribution for a configuration of the three-state alternate walk and the four-state walk resulting in localization.
Though both the probability distributions after 30 steps of the walk show localization manifesting from the evolution, the distributions are not identical each other.

To make a comparison between these two walks, we calculate an entanglement measure called the negativity~\cite{LeeChiOhKim2003}, defined by 
\begin{equation}
N(\rho) = \frac{\norm{\rho^{T_b} }-1}{d-1},
\end{equation}
where $\rho^{T_b}$ is the partial transpose of a state $\rho$ in $d_1\otimes d_2$ $(d_1 \leq d_2)$  quantum systems and $\norm{\,\cdot\,}$ is the trace norm.
This will bound the maximum value of the entanglement measure to $1$ for the system of all dimensions. 

\begin{figure}[h]
 \begin{center}
  \begin{minipage}{60mm}
   \includegraphics[scale=0.3]{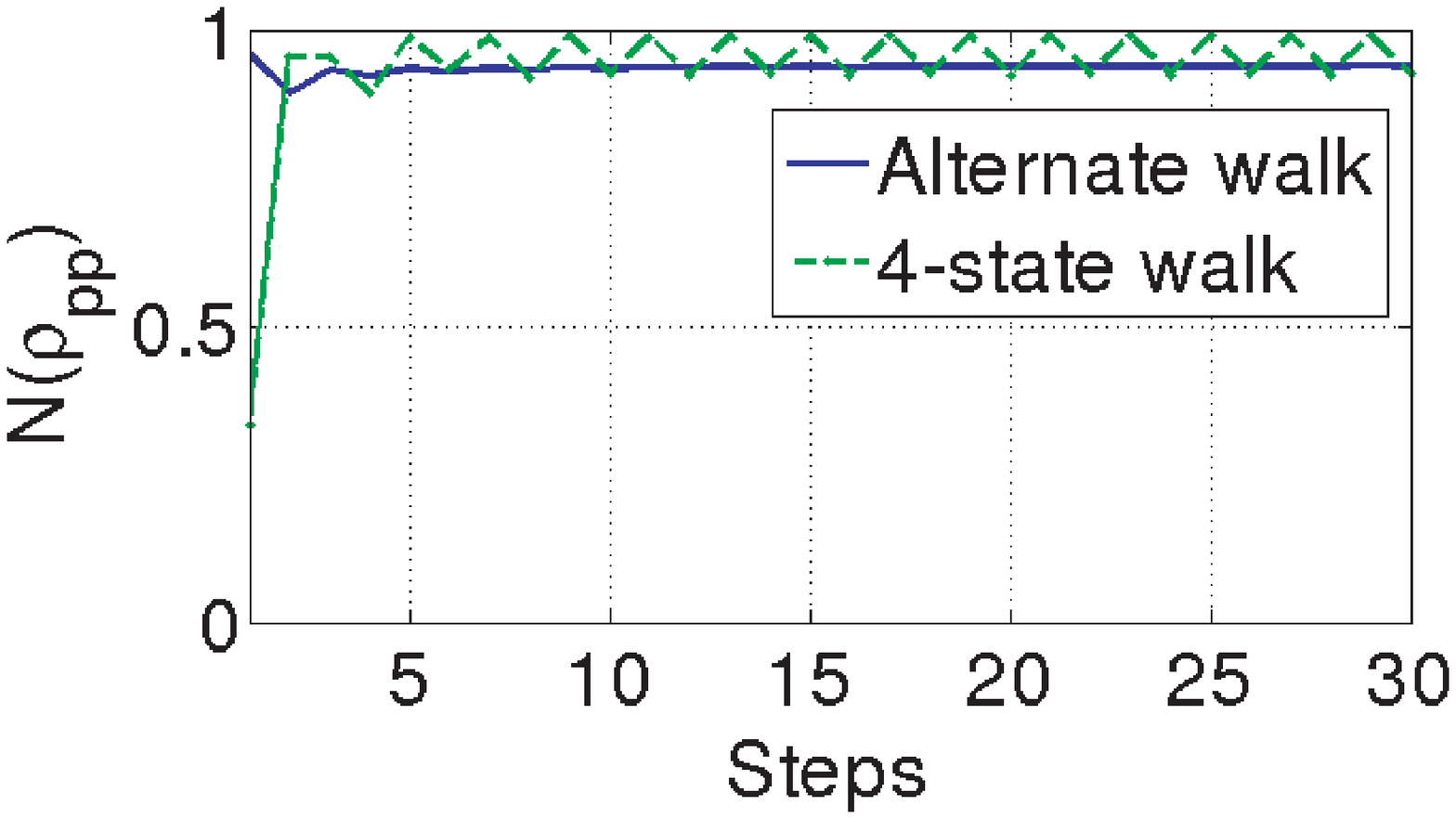}\\[2mm]
   {(a) Particle-Position Entanglement}
  \end{minipage}
  \begin{minipage}{60mm}
   \includegraphics[scale=0.3]{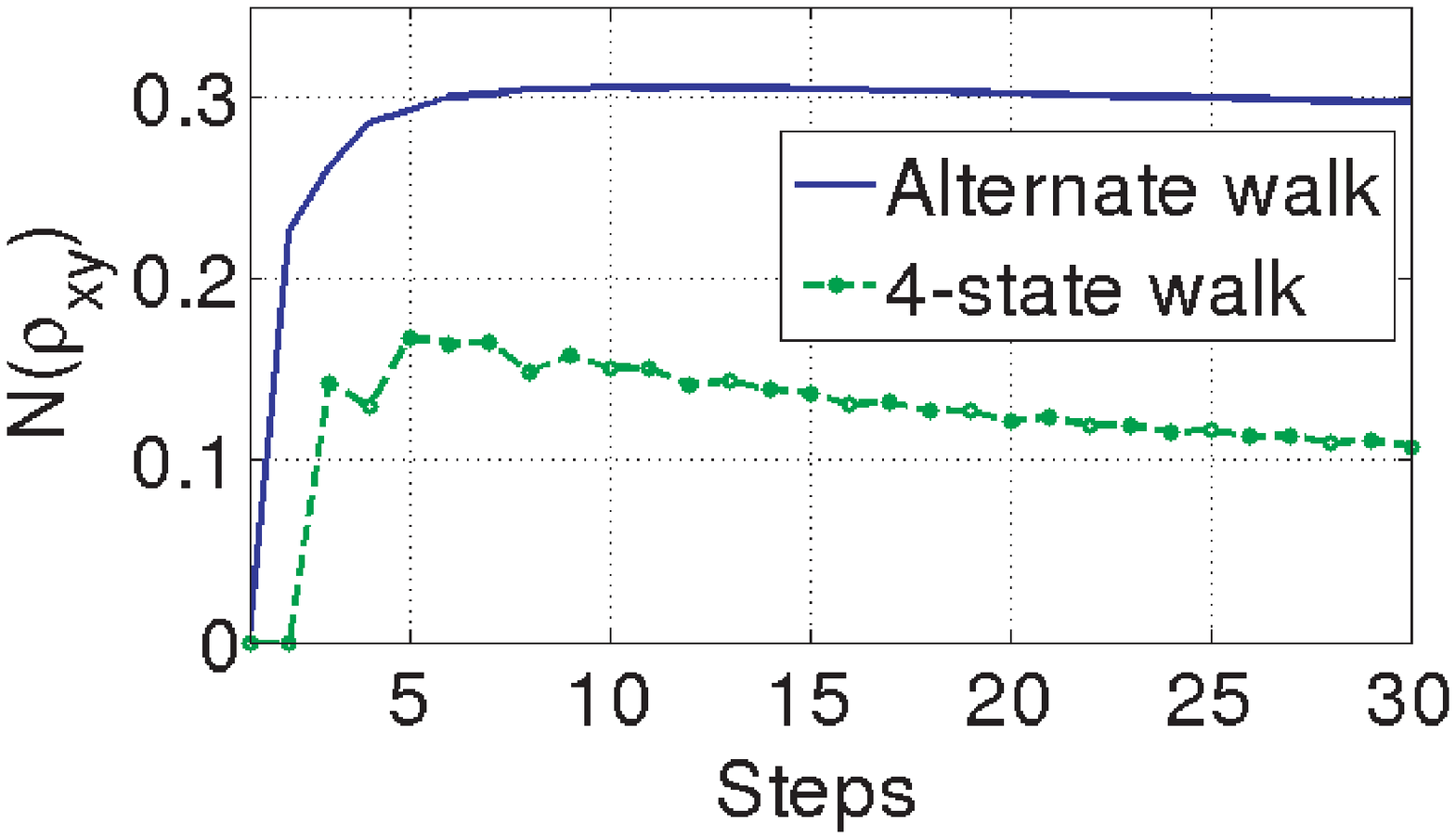}\\[2mm]
   {(b) Spatial entanglement ($X_t=Y_t$)}
   \end{minipage}
  \caption{Negativity as a measure of entanglement for the three-state alternate walk and the four-state walk as a function of steps (time). (a) negativity between particle and position space (b) negativity between the two spatial dimensions ($x-y$).}
  \label{fig:entang} 
 \end{center}
\end{figure}

In Fig.~\ref{fig:entang}(a) we present the negativity between the coin and the position space ($N(\rho_{pp})$) where $\rho_{pp} = \ket{\Psi_t}\bra{\Psi_t}$.
We can see that for both, the three-state and the four-state DTQW the value of negativity is nearly same (close to one) and only for the four-state walk we see the oscillations around the mean value. However, to compare the entanglement generated in different systems we need to consider the system of the same dimension.
Therefore, as mentioned earlier in this section, by tracing out the coin space from the density operator $\rho_{pp}$ of both, the three-state and the four-state DTQW we will be left with the reduced density operator $\rho_{xy}$ of the same dimension and that can be used to calculate the entanglement between the two spatial dimensions.
In Fig.~\ref{fig:entang}(b) the negativity between the two spatial dimensions ($N(\rho_{xy})$) is shown.
We can see that the value of the negativity is very large for evolution using the three-state alternate walk compared to the four-state walk.
This result is in consistency with the results showing a higher spatial entanglement for a two-state alternate DTQW compared to the four-state walk~\cite{DiMcBusch2011, ChandrashekarBusch2013}.
This in general shows that the DTQW on a smaller Hilbert space results in a higher spatial entanglement in the system.
Though it might not be of any immediate physical significance, it has a potential to be a resource when the spatial entanglement is effectively tapped as a resource for quantum information protocols. 

Before we conclude, we will look into the physical realizability of the three-state alternate walk.
Compared to the four- and higher level systems, the three-state system is practically accessible to experimentalists.
Atomic systems in the form of three level systems have played a significant role in demonstrating various interesting coherent phenomenon and generate diverse quantum effects for example, two-photon coherence~\cite{GrischkowskyLoyLiao1975}, coherent multi-level photon ionization~\cite{ThieleGoodmanStone1980, AdachiNikiIzawaNakaiYamanaka1991}, and STIRAP~\cite{BergmannTheuerShore1998}. Recent experimental advancements have been able to demonstrate sufficient control over the three different types of three-level systems know as, $\Lambda$, cascade, and $V$ systems.
Combing the expertise developed in handling the superposition state of the three-level atomic system with experimental expertise demonstrated in implementing the two-state DTQW using atoms on optical lattice~\cite{KarskiForsterChoiSteffenAltMeschedeWidera2009}, a three-state alternate walk is not far off from being a reality.

\section{Discussion and summary}
\label{conc}

We studied a three-state alternate DTQW starting from the origin and obtained two limit theorems.
The walker, which has three coin-states at each position, moves on a two-dimensional position space $\mathbb{Z}^2$ (square lattice) by alternately repeating a walk in $x$-direction followed by a walk in $y$-direction.
The coin-flip operation for the evolution of the walk in each dimension was given by a parameterized unitary matrix which includes a Grover coin.
One of the two limit theorems was the limit of a return probability defined by the probability that the walker returns to the starting point, and the other was a convergence in distribution for the position of the walker on a rescaled space by time.
The return probability can be positive depending on both the initial condition at the origin and the coin-flip operator.
The limit distribution in the convergence on the rescaled space has both a Dirac $\delta$-function at the origin and a continuous function with a compact support which is described by an ellipse.
From these results, we see that the three-state alternate DTQW can localize around the origin.
Using negativity as a measure of entanglement, we also showed that the entanglement generated between the two spatial dimensions using the three-state DTQW is higher than the spatial entanglement generated using the four-state DTQW.

Although we computed just the return probability at a long-time limit, we can also get the long-time limit of the probability $\mathbb{P}[(X_t,Y_t)=(x,y)]$ for any $x,y\in\mathbb{Z}$ according to Eqs.~\eqref{eq:1st-state},\,\eqref{eq:2nd-state}, and \eqref{eq:3rd-state}.
It is, however, hard to calculate the limit due to the function $F(x,y)$ which is a single variable integral in Eq.~\eqref{eq:single_value_integral}.
To know the behavior of the walker after many steps at any position besides the origin, it would be an interesting future problem to compute the integral $F(x,y)$ and the limit $\lim_{t\to\infty}\mathbb{P}[(X_t,Y_t)=(x,y)]$. 




\bc
{\bf ACKNOWLEDGEMENTS}
\ec
TM acknowledges support from the Japan Society for the Promotion of Science.

\end{document}